\documentclass[11pt]{amsart}

\usepackage{amsgen,amsthm, amsmath,amstext,amsbsy,amsopn,amsfonts,amssymb}

\newcommand\la{\langle}
\newcommand\ra{\rangle}

\newcommand\bb{{\mathfrak b}}
\newcommand\dd{{\mathfrak d}}

\newcommand\ggo{{\mathfrak g}}
\newcommand\hh{{\mathfrak h}}

\newcommand\pp{{\mathfrak p}}

\newcommand\sso{{\mathfrak s}}
\newcommand\vv{{\mathfrak v}}
\newcommand\uu{{\mathfrak u}}

\newcommand\HH{\mathbb H}

\newcommand\RR{\mathbb R}
\newcommand\ZZ{\mathbb Z}

\newcommand\End{\operatorname{End}}

\newcommand\ad{\operatorname{ad}}
\newcommand\Ad{\operatorname{Ad}}

\theoremstyle{plain}

\newtheorem{thm}{Theorem}[section]

\newtheorem{prop}[thm]{Proposition}
\newtheorem{cor}[thm]{Corollary}

\theoremstyle{definition}
\newtheorem{defn}[thm]{Definition}
\newtheorem{remark}[thm]{Remark}
\newtheorem{example}[thm]{Example}

\begin{document}

\title[Small oscillations and  the Heisenberg Lie algebra]
{Small oscillations  and the Heisenberg Lie algebra} \footnote{}

\author{Gabriela Ovando}

\thanks{{\it (2000) Mathematics Subject Classification}: 70G65, 70H05, 70H06, 22E70, 22E25 }


\address{CIEM - Facultad de Matem\'atica, Astronom\'\i 
a y F\'\i sica,
Universidad Nacional de C\'or\-do\-ba, C\'or\-do\-ba~5000, Argentina}
\email{ovando@mate.uncor.edu}


\begin{abstract} The Adler Kostant Symes [A-K-S] scheme is used to describe mechanical systems for quadratic Hamiltonians of $\RR^{2n}$ on coadjoint orbits of the Heisenberg Lie group. 
The coadjoint orbits are realized in a solvable Lie algebra $\ggo$ that admits an ad-invariant metric. Its quadratic induces the Hamiltonian on the orbits, whose  Hamiltonian system is equivalent to that one on $\RR^{2n}$. This  system is a Lax pair equation whose solution can be computed with help of the Adjoint representation. 
For a certain class of functions, the Poisson commutativity on the coadjoint orbits in  $\ggo$ is related to the commutativity of a family of derivations  of the 2n+1-dimensional Heisenberg Lie algebra $\hh_n$. Therefore the complete integrability is related to the existence of an n-dimensional abelian subalgebra of certain derivations in $\hh_n$. For instance, the  motion of n-uncoupled harmonic oscillators near an equilibrium position can be described with this setting.
\end{abstract}

\maketitle

 \noindent\section{Introduction}
 
 A quadratic  Hamiltonian is a function of the form
\begin{equation}\label{ha1}
H(x)=\frac12 (Ax, x)
\end{equation}
 where $x=(q_1, \hdots, q_n, p_1, \hdots, p_n)$ is a vector in $\RR^{2n}$ written
  in a symplectic  basis and $A$ is a symmetric linear operator with respect to the canonical inner product
 $( , )$. The Hamiltonian  equation  have the form
 \begin{equation}\label{ham1}
 x'=JA x , \qquad\mbox{ where } J = \left( \begin{matrix} 0 & -Id \\ Id & 0
 \end{matrix} \right)
 \end{equation}
 where $Id$ denotes the identity.
 
 In particular the motion of n uncoupled harmonic oscillators can be approximated
 by a quadratic Hamiltonian $H$ as in (\ref{ha1})  with $A=Id$,  explicitely in coordinates 
 $
 H=\frac 12 \sum_i (p_i^2+q_i^2)
 $
 where $q_i$ denote the position coordinates and $p_i=\dot{q_i}$ are the 
 canonical momentum coordinates. Then (\ref{ham1}) yields the  equations of motion, which  predict the position and the velocity  at any time if initial conditions $q_i(t_0)$, $p_i(t_0)=\dot{q_i}(t_0)$ are 
known.

 In quantum mechanics a good approach to the simple harmonic oscillator is through the
 Heisenberg Lie algebra. In dimension three this is the Lie algebra generated by the
 position operator $Q$ = {\em multiplication by x}, the momentum operator $P = -i \frac
 {d}{dx}$ and $1$ with the only non trivial commutation relation
 $$[Q,P]=1$$
 These operators evolve according to the Heisenberg equations
 $$\frac{dP}{dt}=-Q \qquad \qquad \frac{dQ}{dt}= P$$
 
 In this work we shall show that the 2n+1-dimensional Heisenberg Lie algebra  $\hh_n$
 allows an approach  to systems associated to 
 quadratic Hamiltonians of the form (\ref{ha1}) in classical mechanics. To this end we shall make use of Lie theory. Lie theory
 was successful when studying some mechanical systems such as the rigid body or the generalized 
 Toda lattice \cite{A} \cite{Ko2} \cite{Sy} \cite{R2}. In this
 setting ad-invariant functions play an important role. On the one hand their Hamiltonian systems becomes a Lax 
 equation and on the other hand they are in involution on the orbits. Whenever studying Poisson commuting conditions the ad-invariance property can be replaced by a weaker one as in
 \cite{R1}. 
 In the framework of this theory what we
 need is a Lie algebra with an ad-invariant metric, a splitting of this Lie algebra into
 a direct sum as vector subspaces of two subalgebras and a given function. These algebraic tools were used with semisimple Lie algebras, where the Killing form is the natural candidate for the ad-invariant
 metric. 
 
However there are more Lie algebras admitting an ad-invariant
 metric. For our purposes we are interested in the solvable ones. They can be constructed by a double extension procedure, whose more simple application follows from $\RR^m$. We get a solvable Lie algebra $\ggo$, that results a semidirect extension of the 2n+1-dimensional Heisenberg Lie algebra $\hh_n$ and that can be endowed with an ad-invariant metric which is an extension of the bilinear form on $\RR^{2n}$ given by $b(x, y)= (Ax, y)$ if $(, )$ is the canonical inner product on $\RR^{2n}$ and $A$ is a non singular symmetric transformation. The set of symmetric maps $A$ is in a bijective correspondence with the derivations of the Heisenberg Lie algebra acting trivially on the center, a set denoted with $\dd$. Any such derivation allows a semidirect extension $\ggo$ that admits an ad-invariant metric. 
 
 The Lie algebra $\ggo$ splits naturally as  a direct sum of vector spaces of a one dimensional Lie algebra and the Heisenberg ideal. Looking the coadjoint orbits of the Heisenberg Lie group on $\ggo$ via the metric, one gets  Hamiltonian systems on these orbits. In particular for the restriction of the quadratic corresponding to the ad-invariant metric we obtain  a Hamiltonian system equivalent to (\ref{ha1}). Since the considered function is ad-invariant the system  becomes a Lax equation, whose solution can be computed with the Adjoint representation. 
 
 As example we work out the linear equation of motion of n-uncoupled harmonic oscillators. The Lie algebra $\ggo$ is known as a oscillator Lie algebra and the corresponding ad-invariant metric is Lorentzian. This Lie algebra belongs to the exclusive class of Lie algebras admitting an ad-invariant Lorentzian metric. 
The quadratic for this metric induces the Hamiltonian system on the orbits, whose solutions   are bounded. Furthermore it is proved that the Hamiltonian  is completely integrable on all maximal orbits. We notice that the functions in involution we are making use, are not ad-invariant and they do not satisfy the involution conditions of \cite{R1}.
 
 Going back to the general case  we study involution conditions on the orbits for a class  of quadratic functions which are not ad-invariant in $\ggo$ in general. They are functions of the form (\ref{ha1}) extended to $\ggo$.  The Poisson commutativity conditions we get for these functions can be read off  in the Lie algebra of derivations of $\hh_n$. If the centralizer of $JA$ in $\dd$ (J as in (\ref{ham1}) and $A$ as above)  contains an n-dimensional abelian subalgebra, then the function (\ref{ha1}) is completely integrable on $\RR^{2n}$ whose related one on $\ggo$ is completely integrable on all orbits. This reduces the complete integrability of these systems to algebraic conditions on $sp(n)$, the Lie algebra of derivations of $\hh_n$ acting trivially on the center. Applying results of Lie Theory, one can see that many of these Hamiltonians are completely integrable. One needs to study the abelian subalgebras of $sp(n)$.  In  particular for the case of the motion of n-uncoupled harmonic oscillators we need a abelian subalgebra in the Lie algebra of isometries of the Heisenberg Lie group $\HH_n$, endowed with its canonical inner product.

 \section{Preliminaries}

Let $G$ be a Lie group with Lie algebra $\ggo$ and  exponential map 
$\exp:\ggo \to G$.  

Let us introduce a kind of Lie algebras we shall work with.
For each integer $i\geq 1$ define $\ggo^i= [\ggo,
\ggo^{i-1}]$, where $\ggo^0=\ggo$. The Lie algebra $\ggo$ is {\em nilpotent} if
$\ggo^i=0$ for some positive integer $i$. It is said {\em k-step nilpotent} if $\ggo^k=0$
but $\ggo^{k-1}\neq 0$. If $G$ is the unique simply connected nilpotent Lie
group corresponding to a given nilpotent Lie algebra $\ggo$, then the
exponential map $\exp: \ggo \to G$ is a diffeomorphism \cite{Ra}.

\begin{example} \label{hn} The Heisenberg Lie algebra is an example of a 2-step nilpotent
 Lie algebra,
whose Lie group is called the Heisenberg Lie group $\HH_n$. It can be  constructed on 
$\RR^n \times \RR^n
\times \RR$ with the canonical topology and  with the product
\begin{equation}
(x,y,x_0)(x',y',x_0')=(x+x',y+y',x_0+x_0'+\frac12(x \cdot y'-y \cdot x')),
\end{equation}
 where $x,x',y,y'\in \RR^n$ and $x_0,x_0'\in \RR$.
The identity element coincides with the origin $0=(0,0,0)$.
$\HH_n$ is non commutative and its center $\ZZ(\HH^n)$ is
the set of elements $(0,0,x_0)$ with $x_0\in \RR$. Thus
$\HH^n/{Z(\HH^n)} \simeq \RR^{2n}$
in the sense that every class is determined by the first two
components $(x,y)$ and the product in the quotient group coincides
with the sum on $\RR^{2n}$.

The left invariant vector fields at a point  on $\RR^{2n+1}$ are
$$X_j=\frac{\partial}{\partial x_j} -\frac{y_j}2
\frac{\partial}{\partial x_0}, \qquad Y_j=\frac{\partial}{\partial
y_j} +\frac{x_j}2 \frac{\partial}{\partial x_0},\qquad X_0=
\frac{\partial}{\partial x_0}.$$ It is easy to verify the Lie
bracket relations: $$[X_i,Y_j]=\delta_{ij}X_0.$$ Let $\hh_n$ be the
Heisenberg Lie algebra of dimension 2n+1, that is the Lie algebra
of left invariant vector fields on $\HH_n$, which coincides with
$T_0(\HH^n)$. Define a definite metric on $T(\HH_n)$ so that the
vectors $X_i, Y_j, X_0$ are orthonormal for all i,j=1,$\hdots$, n. Then this metric is left
invariant and can be transported to the quotient space
$\HH_n/Z(\HH_n)$. The induced metric coincides with the canonical
one on $\RR^{2n}$. 

In the following section we shall see another construction
of the Heisenberg Lie algebra with help of the Poisson bracket on $\RR^{2n}$.
\end{example}

Another class of Lie algebras (groups) is constitued by  the solvable ones.
Define ideals $D^i(\ggo)$ in $\ggo$  by
$D^i(\ggo)=[D^{i-1}(\ggo), D^{i-1}(\ggo)]$ where $D^0(\ggo)=\ggo$. The {\em
solvable} Lie algebras are those for which there exists an integer $k$ such that
$D^k(\ggo)=0$. In solvable Lie algebras the commutator $C(\ggo)=[\ggo, \ggo]$ is
a nilpotent ideal.

 Let $M$ be a smooth manifold and $\phi:G \times M \to M$ be a smooth action of $G$ on $M$. The
 vector fields on $M$
 $$\tilde{X}(m)=\frac{d}{dt}_{|_{t=0}} \phi(\exp tX,m) \qquad m \in M, \quad X\in \ggo,
 \quad \, t\in \RR$$
 will denote the infinitesimal generators of this action. If $G\cdot
 m=\{\phi(g,m),\, g
 \in G\}$ denotes the $G$-orbit through $m\in M$  its tangent space is the set 
 $$T_m(G\cdot m)=\{ \tilde{X}(m)\, /\, X\in \ggo\}.$$
 
 Here we also make use of the notation $g\cdot m =\phi(g,m)$. 
 The following actions are important in our setting:
 
 - the adjoint action $\Ad:G\times \ggo \to \ggo$  whose infinitesimal generators are
 $\tilde{X}=\ad_X$, where $\ad_X Y=[X,Y]$ denotes the Lie bracket of $X,Y \in \ggo$;
 
 - the coadjoint action of $G$ on $\ggo^{\ast}$ is the dual of the adjoint action and it
 is given by $g\to \Ad^{\ast}(g^{-1})$, for $g\in G$, whose infinitesimal generator is
 $\tilde{X}=-\ad_{X}^{\ast}$.
 
 The coadjoint orbits are examples of symplectic manifolds. Recall that they are endowed
 with the Kirillov-Kostant-Souriau symplectic structure given by:
 $$\omega_{\beta}(\tilde{X},\tilde{Y})=-\beta ([X,Y]), \qquad \beta \in G\cdot \mu.$$
 Assume now that $\ggo$ has an ad-invariant metric  $\la ,
\ra: \ggo\times \ggo \to \RR$; , that is, $\la , \ra$ is  non-degenerate symmetric bilinear form for which the adjoint representation is skew symmetric. This gives rise to a bi-invariant pseudo Riemannian metric on the Lie group $G$ with Lie algebra $\ggo$; bi-invariant means that the maps $\Ad(g)$ are isometries
for all $g\in G$. Then  $\la, \ra$ induces a diffeomorphism between the adjoint orbit $G\cdot X$ and the
coadjoint orbit $G\cdot \ell_X$ where $\ell_X (Y)=\la X, Y\ra$.

Recall that given a metric $\la , \ra$ on $\ggo$  the gradient of a function $f:\ggo  \to \RR$ at the vector $X\in \ggo$ 
is defined  by
 $$\la \nabla f(X), Y\ra = df_X(Y)\qquad \qquad y\in \ggo. $$

Suppose that the Lie algebra $\ggo$ admits a splitting
$$\ggo  = \ggo_+ \oplus \ggo_-$$ as a direct sum of linear subspaces, where $\ggo_+$,
$\ggo_-$ are subalgebras of $\ggo$. Then the Lie algebra $\ggo$ also splits as
$\ggo=\ggo_+^{\perp} \oplus \ggo_-^{\perp}$, where $\ggo_{\pm}^{\perp}$ is isomorphic as
vector spaces  (via $\la , \ra$) to $\ggo_{\mp}^{\ast}$. Let $G_-$
denote a subgroup of $G$ with Lie algebra $\ggo_-$. Then the coadjoint action of
$G_-$ on $\ggo_-^{\ast}$ induces an action of $G_-$ on $\ggo_+^{\perp}$:
$$g_- \cdot X= \pi_{\ggo_+^{\perp}}(\Ad(g_-)X ) \quad g_-\in G_-, \quad X \in
\ggo_+^{\perp},$$
 where $\pi_{\ggo_+^{\perp}}$ denotes the projection of $\ggo$ on
 $\ggo_+^{\perp}$. Thus the
 infinitesimal generator corresponding to $X_-\in \ggo_-$ is 
 $$\tilde{X_-}(Y)=\pi_{\ggo_+^{\perp}}([X_-, Y]) \qquad Y\in \ggo_+^{\perp}.$$
 The orbit $G_-\cdot Y$ becomes a symplectic manifold with the
 symplectic structure  given by
 $$\omega_X(\tilde{U_-}, \tilde{V_-})=\la X, [U_-,V_-]\ra \qquad
 \mbox{ for } U_-, V_- \in \ggo_-, X\in G_- \cdot Y$$
 which is induced by the Kostant-Kirillov-Souriau symplectic form on
 the coadjoint orbits in $\ggo_-^{\ast}$.

 Consider the restriction of the function $f:\ggo \to \RR$ to an orbit 
$G_-\cdot X 
 :=\mathcal M\subset \ggo_+^{\perp}$. Then the Hamiltonian vector field 
of
 $H=f_{|_{\mathcal M}}$ is the infinitesimal generator corresponding to $-\nabla f_-$ , that is  \begin{equation}\label{e3}
 X_H(Y)=-\pi_{\ggo_+^{\perp}}([\nabla f_-(Y),Y])\end{equation}
 where  $Z_{\pm}$ denotes the projection of $Z \in \ggo$ with respect 
to the
decomposition $\ggo=\ggo_+\oplus \ggo_-$. In fact for $Y\in
 \ggo_+^{\perp}$, $V_-\in \ggo_-$ we have 
$$\begin{array}{rcl}
\omega_Y(\tilde{V_-}, X_H) & = & dH_Y(\tilde{V_-}) = \la \nabla f(Y),
\pi_{\ggo_+^{\perp}}([V_-,Y])\ra= \la \nabla f_-(Y),[V_-,Y]\ra \\
 & = & \la Y,[\nabla f_-(Y),V_-]\ra =  \omega_Y(\tilde{\nabla
   f_-(Y)},\tilde{V_-}).
\end{array}$$
 Since $\omega$ is non degenerate, one gets (\ref{e3}). Therefore the 
Hamiltonian equation for $x:\RR \to \ggo$ 
 follows
 \begin{equation}\label{e4}
 x'(t)=-\pi_{\ggo_+^{\perp}}([\nabla f_-(x),x]).\end{equation}
In particular if $f$ is ad-invariant then $0 = [\nabla f(Y),Y]=
[\nabla f_-(Y),Y]+
[\nabla f_+(Y),Y]$. Since the metric is ad-invariant
$[\ggo_+, \ggo_+^{\perp}]\subset \ggo_+^{\perp}$ and thus equation 
(\ref{e4}) takes the form
\begin{equation}\label{e5}
x'(t)=[\nabla f_+(x),x]=[x,\nabla f_-(x)],\end{equation}
hence (\ref{e4}) becomes a  Lax equation, that is, it can be written as $x'=[P(x),x]$.

If we assume now that the multiplication map $G_+\times G_- \to G$,
$(g_+, g_-) \to g_+g_-$,
is a diffeomorphism, then the initial value problem
\begin{equation}\label{e6}
\left\{ \begin{array}{rcl}
\frac{dx}{dt} & = & [\nabla f_+(x),x] \\
x(0) &  = & x_0
\end{array}
\right.
\end{equation}
can be solved by factorization. In fact if $\exp t\nabla 
f(x_0)=g_+(t)g_-(t)$, then
$x(t)=\Ad(g_+(t))x_0$ is the solution of (\ref{e6}).

{\sc Remark.} If the multiplication map $G_+ \times G_- \to G$ is a 
bijection
 onto an
open subset of $G$, then equation (\ref{e4})  has a local solution in 
an interval
$(-\varepsilon, \varepsilon)$ for some $\varepsilon >0$.

Recall that the Poisson bracket on $C^{\infty}(\ggo)$ is given by
$$\{f,h\}(X)=\la X, [\nabla f(X), \nabla h(X)]\ra$$
which is the Poisson bracket associated to the symplectic form on the 
adjoint orbits (the structure is induced via the metric from  the
 coadjoint orbits).

A first step in the construction of action angle variables is to search for functions
which Poisson commute. 
 The  Adler-Kostant-Symes  Theorem shows a way to get functions in
 involution on the orbits $\mathcal M$.  We shall formulate it  in its classical Lie algebra
 setting.

 \begin{thm}\label{AKS1} Let $\ggo$ be a Lie algebra
 with an ad-invariant metric $\la , \ra$. Assume $\ggo_-, \ggo_+$  are Lie subalgebras
 such that $\ggo=\ggo_-\oplus \ggo_+$ as direct sum of vector  subspaces. Then any pair of
 ad-invariant functions on $\ggo$ Poisson commute on $\ggo_+^{\perp}$ (resp. on
 $\ggo_-^{\perp}$).
 \end{thm}
 Sometimes the ad-invariant condition is too strong, so the following version of the
 previous Theorem given by Ratiu \cite{R1} ask for a weaker condition.

 \begin{thm}\label{AKS2} Let $\ggo$ be a Lie algebra carrying  an ad-invariant metric $\la , \ra$. Assume it admits a splitting into a direct sum as vector spaces  $\ggo=\ggo_+\oplus  \ggo_-$, where $\ggo_+$ is an ideal
  and $\ggo_-$ is is a Lie subalgebra. If $f,h$ are smooth Poisson commuting functions on
 $\ggo$, then the restrictions of $f$ and $h$ to $\ggo_+^{\perp}$ are in involution in
 the Poisson structure of $\ggo_+^{\perp}$.
 \end{thm}
\begin{remark} This theorem was used in \cite{R2} to prove the involution of the Manakov
integrals for the free n-dimensional rigid body motion.
\end{remark}

\subsection{The motion of n  Harmonic oscillators}
We shall apply the algebraic scheme of the previous section to the motion of $n$ uncoupled harmonic oscillators. This will be done with a solvable Lie algebra $\ggo$ known as a  oscillator Lie algebra, which can be endowed with an ad-invariant metric. This Lie algebra admits a splitting into a direct sum as vector spaces of two subalgebras. One of them is the Heisenberg Lie algebra, whose corresponding Lie group acts on the coadjoint orbits,  included on $\ggo$ via the ad-invariant metric on $\ggo$. We choose a certain function and we realize the corresponding Hamiltonian system on the coadjoint orbits. The system is clearly equivalent to the linear one that approximates the motion of n uncoupled harmonic oscillators. Moreover we can show the complete integrability of the Hamiltonian on all maximal orbits.

\vspace{6pt}
  
The motion of n uncoupled harmonic oscillators can be approximated
 by a quadratic Hamiltonian $H$ as in (\ref{ha1})  with $A=Id$,  that is
 \begin{equation}\label{e1}
 H=\frac 12 \sum_i (p_i^2+q_i^2)
 \end{equation}
 where $q_i$ denote the position coordinates and $p_i=\dot{q_i}$ are the 
 canonical momentum coordinates. Then (\ref{ham1}) yields the
following  equation of motion
 \begin{equation}\label{hamho}
 \begin{array}{rcl}
 \frac{dq_i}{dt} & = & p_i \\ \\
 \frac{dp_i}{dt} & = & -q_i
 \end{array} 
 \end{equation}
  These equations predict the position and the velocity  at
any time if initial conditions $q_i(t_0)$, $p_i(t_0)=\dot{q_i}(t_0)$ are 
known.
  The phase space in this
case is $\RR^{2n}$, which is a symplectic manifold with the canonical
structure given by
$$ \omega= \sum_i dq_i \wedge dp_i.$$
This has an associated Poisson structure, which for smooth functions
$f, g $ on $\RR^{2n}$ is  defined by
\begin{equation}\label{pb}
\{f, g \} = (\nabla f, J \nabla g)=  \sum_i \frac{\partial f}{\partial q_i}  \frac{\partial
 g}{\partial p_i}-
 \frac{\partial f}{\partial p_i} \frac{\partial g}{\partial q_i}.
\end{equation}
With respect to this Lie bracket $\{ , \}$ the subspace over $\RR$
generated by the functions $H=\frac12\sum_i(q_i^2+p_i^2)$, the coordinates 
$q_i$, $p_i$, and $1$
form a solvable Lie algebra of dimension 2n+2, which is a semidirect extension
of the Heisenberg Lie algebra spanned by the functions $q_i,p_i,1$ i=1, $\hdots$,n. In fact 
they obey
the following non trivial rules
$$\{q_i, p_i \} =1 \qquad \{H, q_i \} =-p_i \qquad \{H, p_i\} =q_i.$$
In order to simplify notations let us rename these elements
identifying $X_{n+1}$ with $H$, $X_i$ with $q_i$, $Y_i$ with $p_i$ and $X_0$
with the constant function 1 and set $\ggo$ denotes the Lie algebra
generated by these vectors with the Lie bracket $[,]$ derived from the
Poisson structure. This Lie algebra is known as a {\em oscillator} Lie algebra.

Consider the splitting of $\ggo$ into a vector space direct sum
  $\ggo = \ggo_+ \oplus \ggo_-$, where $\ggo_{\pm}$ denote the Lie 
subalgebras
\begin{equation}
\label{deco} \ggo _- =  span\{X_0, X_i, Y_j\}_{i,j=1, \hdots n},\qquad 
 \qquad  \ggo_+ = \RR{X_{n+1}}.
\end{equation}

Notice that $\ggo_-$ is isomorphic to the 2n+1-dimensional Heisenberg Lie
 algebra  we denote $\hh_n$.

The quadratic form on $\ggo$ which for
 $X =x_0(X) X_0  + \sum_i (x_i(X) X_i + y_i(X) Y_i) + x_{n+1}(X)X_{n+1}$ is given by
$$f(X)=  \frac12  \sum_i(x_i^2+ y_i^2)+ x_0x_{n+1}$$
induces an ad-invariant metric on $\ggo$ denoted by $\la , \ra$. 
Canonical computations show that the gradient of $f$ at a point $X$ is  $$ \nabla f(X) = X.$$

The restriction of the quadratic form to $\vv:=span\{X_i, Y_j\}$ i, j=1, $\hdots $,
n, coincides with the canonical one on $\RR^{2n}\simeq \vv$.
 In other words the Lie algebra
$\ggo$ is the double extension of $\RR^{2n}$ with the canonical metric by the skew symmetric
linear map which acts on $\vv$ as the restriction of
$\ad(X_{n+1})$ to this space (see for instance \cite{MR} for the double 
extension procedure).

 The  metric induces a decomposition of  the Lie
 algebra $\ggo$
 into a  vector subspace direct sum of
$\ggo_+^{\perp}$ and $\ggo_-^{\perp}$ where $$ \ggo_-^{\perp} =
span\{X_0\} \qquad \qquad \ggo_+^{\perp} = \RR X_{n+1} \oplus \,span\{X_i,
Y_j\}_{i,j=1, \hdots, n},$$
and it also induces linear isomorphisms
$\ggo_{\pm}^{\ast}\simeq\ggo^{\perp}_{\mp}$. Let $G$ denote a Lie
group with Lie algebra $\ggo$ and $G_{\pm}\subset G$ is a Lie subgroup
whose Lie algebra is $\ggo_{\pm}$. Hence the  Lie subgroup
$G_-$ acts on $\ggo_+^{\perp}$ by the ``coadjoint'' representation; indeed in 
terms of
$U_-\in \ggo_-$ and $V\in \ggo_+^{\perp}$ 
the infinitesimal action of $\ggo_-$ on $\ggo_+^{\perp}$ is
\begin{equation}\label{m2}
\begin{array}{rcl}
\ad^{\ast}_{U_-} V  & =  &  x_{n+1}(V) \sum_i(y_i(U)
X_i - x_i(U) Y_i)
\end{array}
\end{equation}
It is not difficult to see that the orbits are 2n-dimensional
if $x_{n+1}(V) \ne 0$ and furthermore $V$ and $W$ belong to the same orbit if and only if
$x_{n+1}(V)=x_{n+1}(W)$, hence the orbits are parametrized by the $x_{n+1}$-coordinate;  
 so we denote them by $\mathcal M_{x_{n+1}}$. They are topologically
like $\RR^{2n}$. In fact $\mathcal M_{x_{n+1}}= G_- \cdot V\simeq
\HH_n/Z(\HH_n)$, where $\HH_n$ denotes the Heisenberg Lie group with center
 $Z(\HH_n)$.
 
 Equipp these  coadjoint orbits  with the canonical symplectic structure. That is 
 $$\omega_Y(\tilde{U_-}, \tilde{V_-})=\la Y, [U_-, V_-]\ra = x_{n+1}(Y) \sum_{i=1}^n (x_i(U_-) y_i(V_-)-x_i(V_-)y_i(U_-)) \quad
U_-, V_- \in \ggo_-.$$
Indeed on  the orbit $\mathcal M_1$  the coordinates $x_i, y_j$, $i,j =1, \hdots n$, are   the canonical 
symplectic coordinates  and one can identify this orbit with $\RR^{2n}$ in a natural way.
 
 Consider $H$, the 
 restriction to a orbit $\mathcal M_{x_{n+1}}$ of  the function $f$. Since $f$ is ad-invariant the
Hamiltonian system of $H=f_{|_{\mathcal M_{x_{n+1}}}}$ reduces to
\begin{equation}\label{osc}\begin{array}{rcl}
\frac{dx}{dt}& = & [x_{n+1}X_{n+1}, x_{\vv}+ x_{n+1} X_{n+1}]\\
 x(0) & = & x^0
\end{array}
\end{equation}
where $x^0= x_{\vv}^0 +x_{n+1}^0 X_{0}$ and $x_{\vv}^0=\sum_i (x_i^0 X_i + y_i^0
Y_i)$.

For $x_{n+1}\equiv x_{n+1}^0\equiv 1$ this system is equivalent to  
(\ref{hamho}).

The trajectories $x(t)$ with coordinates $x_i(t)$, $y_j(t)$, $x_{n+1}^0$  are parametrized 
circles of angular velocity $x_{n+1}^0$,  for all i,j, that is

  $$
\begin{array}{rcl}
x_i(t) & = &  x_i^0 \cos(x_{n+1}^0 t)  + y_i^0 \sin( x_{n+1}^0 t)\\
y_j(t) & = & - x_j^0 \sin(x_{n+1}^0 t) + y_j^0 \cos(x_{n+1}^0 t)\\
x_{n+1}(t) & = &  x_{n+1}^0
\end{array}
$$
 This solution coincides
with that computed in the previous section, when we considered systems 
on coadjoint orbits. In fact it can be written as 
$$x(t)= \Ad(exp\,\, t x_{n+1}^0 X_{n+1}) x^0,$$
and one verifies that the  flow at the point $X^0 \in \ggo_+^{\perp}$ is 
\begin{equation}\label{flow}
\begin{array}{rcl}
\Delta^t(X^0) & = & 
\sum_i [(x_i^0 \cos(x_{n+1}^0 t) + y_i^0 \sin( x_{n+1}^0 t))X_i + 
(-x_i^0 \sin(x_{n+1}^0 t) + \\ \\
& & y_i^0 \cos(x_{n+1}^0 t))Y_i] +   x_{n+1}^0 X_{n+1}
\end{array}
\end{equation}


\vskip .2cm

System (\ref{osc}) is a Lax pair equation $L^{\prime} = [M, L]= ML -
LM$, taking $L$ and $M$ the following matrices for $\omega_i=1$ for all i:
$$ M = \left(
\begin{matrix}
0 & x_{n+1}\omega_1 & 0& 0 &&& &0 & 0\\ -x_{n+1}\omega_1 & 0 & 0 & 0 
&&&&0 &
0 \\ 0 & 0 & 0 & x_{n+1}\omega_2 &&&&0& 0\\ 0& 0 & -x_{n+1}\omega_2&
0&&&&0&0\\ & & & & \ddots & & &\vdots & \vdots\\ & & & & & 0 &
x_{n+1} \omega_n&0 & 0\\ & & & & 0& -x_{n+1}\omega_n& 0 &0& 0\\ 0& 0& 
\hdots
& & & & & 0 & 0\\ 0 & 0 & \hdots & & & &  & 0 & 0
\end{matrix}
\right) $$
$$ L = \left(
\begin{matrix}
0 & x_{n+1}\omega_1 & 0& 0 & & & & &x_1\\ -x_{n+1}\omega_1 & 0 & 0 & 0  
& &
& & & y_1\\ 0 & 0 & 0 & x_{n+1}\omega_2 & & & & & x_2\\ 0& 0 &
-x_{n+1}\omega_2 & 0 & & & & & y_2\\ & & & & \ddots & &  & \vdots &
\vdots\\ & & & & & &  x_{n+1} \omega_n& 0& x_n\\ & & & & &  -x_{n+1}
\omega_n & 0 &0 & y_n\\ -\frac12 y_1& \frac12 x_1&-\frac12 y_2 &\frac12
x_2 &\hdots & -\frac12 y_n& \frac12 x_n &  0 & 0\\ 0& 0 & 0 & 0 &
\hdots  & 0& 0& 0 & 0
\end{matrix}
\right) $$

\

Next we shall prove the complete integrability of the function $H$, that is the restriction to the orbits $\mathcal M_{x_{n+1}}$ of the quadratic form associated to the ad-invariant metric on $\ggo$. We shall show a
set of n-functions that are in involution. However they are not Ad-invariant,
hence they do not satisfy conditions of Theorem (\ref{AKS1}) and Theorem
(\ref{AKS2}) does not hold in this case.

\begin{defn} \label{ci} Recall that a  function $f$ on a 2n-dimensional Poisson manifold
$(M, \{ , \})$ is {\em completely integrable} if there exist $n$ 
functions $f_1, \hdots, f_n$ such that:

i) $\{f, f_i\}= 0$, $\{f_i, f_j\}=0$ for all $1\leq i, j \leq n$,

ii) The differentials $df_1, \hdots, df_n$ are linearly independent on 
a open
set invariant under the flow of $X_f$.
\end{defn}

The Poisson structure on the orbits $\mathcal M_{x_{n+1}}$ is derived from the symplectic structure. Thus for a pair of functions $f, g: \ggo \to \RR$ the Poisson bracket of their respective restrictions $F$, $G$ at any point $X\in \ggo_+^{\perp}$ is given by
$$\{ F, G \} (X)=\la X, [\nabla f_-(X), \nabla g_-(X)]\ra.$$

 Consider $H_i={f_i}_{|_{\mathcal M_{x_{n+1}}}}$ be the 
restrictions to a orbit $\mathcal M_{x_{n+1}}$ of the
functions
\begin{equation}\label{fi}
f_i(X)=  \frac12 (x_i^2 + y_i^2) + x_0 x_{n+1} \quad \mbox{ for }i= 1, \hdots n
\end{equation}
  The
functions $f_i$ are not ad-invariant but their restrictions commute with respect to
the Poisson bracket induced by the Lie-Kirillov-Kostant symplectic
form on the orbit. Moreover we assert

\begin{prop} \label{ci1}
The function $H$ is completely integrable on the orbits $\mathcal
M_{x_{n+1}}$  for all $x_{n+1}\neq 0$. 
\end{prop}
 \begin{proof} To prove this we first need to compute  the gradient of $f_i$  which is 
$\nabla
f_i(X)=   x_i X_i +  y_i Y_i + x_0 X_0 + x_{n+1} X_{n+1}$. 

Since $[\nabla f_i(X), \nabla
f_j(X)]=0$
 for all i,j, then $H_i, H_j$  Poisson commute on the orbits.  In fact for $X\in
 \ggo_+^{\perp}$ 
  $$\{H_i,H_j\}(X)= \la X,
[{\nabla f_i}_-(X), {\nabla f_j}_-(X)]\ra =0$$
where $V_-$ denotes the projection of $V\in \ggo$ with respect to the splitting
$\ggo=\ggo_+ \oplus \ggo_-$.
 One can verify that the differentials $df_i$ are linearly
independent at any point and one proves the Poisson commutation with 
$H$,
 $\{H, H_i\}=0$ for all $i$.
\end{proof}

The set
$$\begin{array}{rcl}
\mathcal N_{x_{n+1}} & = & \cap_{i=1}^n \{X \in \mathcal M_{x_{n+1}} : H_i(X) =
c_i\}\\
& = & \cap_{i=1}^n\{X= \sum_i(x_i X_i + y_i Y_i) +x_{n+1} X_{n+1}\in 
\mathcal M_{x_{n+1}}
: x_i^2+ y_i^2 = c_i\}
\end{array}$$
 is compact and nonempty for $c_i \ge 0$ for all i. Thus the Liouville 
Theorem
applies  and so it is possible to construct action angle coordinates, which can be 
written as $(H_1, \hdots, H_n, \theta_1, \hdots, \theta_n)$ where $\theta_i$
are the angle variables on the torus $\mathcal N_{x_{n+1}}$. In particular the
flow $X_H$ in coordinates $(H_1, \hdots, H_n, \theta_1, \hdots, \theta_n)$ is
linear.

\begin{remark} Let $g:\ggo \to \RR$ be the quadratic polynomial of $\ggo$ given by 
$g(X)=\la A X_{\vv}, X_{\vv}\ra_{\vv}$, 
 where $A$ is  symmetric with respect to $\la , \ra_{\vv}$. Simple computations show that the gradient of $g$ has the form
$ \nabla g(X) = AX_{\vv}$ for $X \in \ggo.$ Let $H={g}_{|_{\mathcal M_{x_{n+1}}}}$ be the restriction of $g$ 
to the orbit, then following  (\ref{e4}) the Hamiltonian system for $H$ is $x'  =  [x_+, A x_{\vv}]= x_{n+1} \ad_{X_{n+1}} A x_{\vv}= x_{n+1}JA x_{\vv}$ where $x$ is a curve on a orbit $x(t) \subset \mathcal
M_{x_{n+1}}\subset \ggo_+^{\perp}$ for all t. This Hamiltonian system is not a Lax pair equation. In the following section we shall consider a realization of this  system like (\ref{ham1}) as a Lax pair equation.
\end{remark} 


\section{Quadratic Hamiltonians and coadjoint orbits} In this section we shall prove that Hamiltonian systems corresponding to quadratic Hamiltonians in $\RR^{2n}$  of the form $H(x) = \frac12(Ax,x)$ where $A$ is a symmetric map, can be described using the  scheme of the previous section on a solvable Lie algebra. First we shall study the construction of solvable Lie algebras admitting an ad-invariant metric. We prove that quadratic Hamiltonians in $\RR^{2n}$ of the form  (\ref{ha1}) can be extended to a quadratic function on a solvable Lie algebra. This is the quadratic corresponding to an ad-invariant metric on the Lie algebra. These Lie algebras are semidirect extensions of the Heinsenberg Lie algebra. Secondly we shall equipp these solvable Lie algebras with an ad-invariant metric, on which we include the coadjoint orbits of the Heisenberg Lie group. On these orbits we realize the Hamiltonian systems corresponding to the quadratic of the ad-invariant metric. In this way the Hamiltonian system becomes a Lax equation, whose solution can be computed with help of the adjoint map in the solvable Lie group. Finally we discuss involution conditions for a class of functions on these codajoint orbits.

\subsection{The structure of solvable Lie algebras admitting an ad-invariant metric} Let $\ggo$ denote a Lie algebra endowed with an ad-invariant metric $\la, \ra$, that is, $\la , \ra:\ggo \times \ggo \to \RR$ is a non degenerate symmetric bilinear form satisfying
 \begin{equation}\label{ad}
 \la [x,y], z \ra + \la y,[x,z]\ra=0 \qquad \text{ for all} x,y, z \in \ggo.
 \end{equation}
 
 Examples of real Lie algebras admitting an ad-invariant metric are the semisimple ones equipped with its Killing form $B$: $B(x,y)=tr(\ad_x \ad_y)$ where $tr$ denotes the trace and $\ad:\ggo \to \End(\ggo)$ is the adjoint representation.
 
 If $G$ is a connected Lie group with Lie algebra $\ggo$ then the pseudo Riemannian metric on $G$ obtained by left tranlations is also right invariant, or equivalently Ad-invariant.
 
 Lie algebras provided with an ad-invariant metric can be obtained in
 the following way.
Let $(\bb, \varphi)$ be a orthogonal Lie algebra and let $S$ be a skew symmetric derivation of $(\bb, \varphi)$.  Consider the vector space direct sum $\RR Z \oplus \bb \oplus \RR T$ and equipp this vector space with the following Lie bracket: 
$$[z_1 Z + B_1+t_1 T, z_2 Z + B_2 + t_2 T]=\varphi(S B_1, B_2) Z + [B_1, B_2]_{\bb} + t_1 S B_2 - t_2 S B_1$$
where $z_i, t_i\in \RR, i=1,2,$ $B_1, B_2 \in \bb.$
 The metric $\la , \ra$ on $\ggo = \RR Z \oplus \bb \oplus \RR T$ obtained as a orthogonal extension of $\varphi$, that is given by $\la , \ra_{\bb \times \bb}= \varphi$ and $\la Z, T\ra=1$ allows to extend $\varphi$ to  an ad-invariant metric on $\ggo$. The Lie algebra $(\ggo, \la , \ra)$ is called the {\it double extension} of $(\bb, \varphi)$ by $(\RR, S)$.

It can be proved that any solvable Lie algebra $\ggo$  endowed with an ad-invariant  metric $\la , \ra$  is a double extension of a solvable Lie algebra with an ad-invariant metric $(\bb, \varphi)$ by  $\RR$ with a  certain skew symmetric derivation $S$ (see \cite{MR} for instance). 

The first examples of this method to get solvable Lie algebras with ad-invariant metrics follow  from $\RR^m$ endowed with a non degenerate symmetric form $b$.  This $b$ can be written as
$$b(X, Y) = (A X, Y) \qquad \mbox{ being } A \mbox{ symmetric with respect to } ( , ),$$
where  $( , )$ is the canonical inner product on $\RR^{m}$.  Moreover $b$ is non degenerate if and only if $A$ is non singular. 

Assume that $m=2n$. Then the quadratic form corresponding to $b$ takes the form (\ref{ha1}). Let $S$ be a linear transformation on $\RR^{2n}$. It is  skew symmetric with respect to $b$, that is  $b(S X, Y) = - b( X, SY)$ if and only if $(AS X, Y) = - (AX, SY)$. 

If we apply the double extension procedure to $(\RR^{2n}, b)$ by $S$ we get the solvable Lie algebra $\ggo = \RR Z \oplus \RR^{2n} \oplus \RR T$ with the Lie bracket
$$[z_1 Z + B_1+t_1 T, z_2 Z + B_2 + t_2 T]=(A S B_1, B_2) Z + t_1 S B_2 - t_2 S B_1.$$
Clearly $\RR Z \oplus \RR^{2n}$ is a nilpotent ideal with a one dimensional commutator, hence $\ggo$ is a  semidirect  extension of a Heisenberg Lie algebra $\hh_{m/2}$. Furthermore $m/2=n$ if and only if $AS$ is non singular. This says that if $A$ is injective,  any non singular skew symmetric map for $b$ gives rise to a derivation of $\hh_n$ acting trivially on the center.

We shall prove a correspondence between the symmetric maps $A$ in $\RR^{2n}$ and the derivations of the Heisenberg Lie algebra acting trivially on the center.

If we fix the inner product on $\hh_n$ defined in Example (\ref{hn}) (denoted $\la , \ra'$) then the Lie bracket on $\hh_n = \RR X_0 \oplus \vv$ where $\RR^{2n}\simeq \vv = span\{ X_i, Y_j\}_{i,j=1, \hdots, n}$ can be expressed as 
$$\la [X, Y], x_0 X_0 \ra' = x_0 \la J X, Y \ra' \quad \mbox{ with } J \mbox{ as in } (\ref{ham1})$$
and notice that $\la , \ra'_{|_{\vv \times\vv}} = ( , )$. 
A derivation $D$ of $\hh_n$ acting trivially on the center must satisfy $[DU, V]= - [U, DV]$ for all $U, V \in \vv$. Equivalently in terms of $\la , \ra'$, we have that a map $D$ in $\hh_n$ is a derivation acting trivially on the center of $\hh_n$ if and only if the restriction of $D$ to $\vv$ (denoted also $D$) satisfies
$$ ( J DU, V ) = - (JU, DV) \qquad \mbox{ for all } U, V \in \vv,$$ 
where we replaced $\la , \ra'_{\vv}$ by $( , )$ since they coincide on $\vv \simeq \RR^{2n}$. Denote by $\dd$  the set of derivations on $\hh_n$ acting trivially on the center of $\hh_n$. 

\begin{thm} \label{bij} There is a bijection between the set of derivations of $\hh_n$ acting trivially on the center and the set $\sso$ of symmetric linear maps on $\RR^{2n}$. Explicitely this correspondence is given by the linear isomorphism $\psi$, which applies $D \in \dd \to JD \in \sso$, where $J$ is the complex structure as in (\ref{ham1}). 

Moreover one can see that if $A$ is a symmetric map on $\RR^{2n}$ then $JA$ is always a skew symmetric map with respect to the bilinear map defined by $b(X, Y) = (AX, Y)$ . 
\end{thm}
\begin{proof} Let $D$ be an element of $\dd$, then $JD$ is symmetric with respect to the canonical inner product on $\RR^{2n}$. In fact $( JD U, V ) = - (JU, DV)$ since $D$ is a derivation of $\hh_n$ and the assertion follows since $J$ is skew symmetric for $(, )$. For the converse let $A$ be a symmetric transformation relative to $( , )$. Define a map $D$ on $\hh_n$ by $D= JA$ on $\vv$ and extend it trivially to the center. It is easy to see that $D$ is a derivation of $\hh_n$. 

To prove the second assertion let $A$ be a symmetric map on $\RR^{2n}$. Then we have $b(JAX, Y) = (AJAX, Y) = -(AX, JA Y) = -b(X, JA Y)$.  
\end{proof} 

Thus for any non singular derivation $D$ of $\dd$ there always exists a symmetric non degenerate  bilinear form  on $\RR^{2n}$ with respect to which $D$ is skew symmetric. This is given by $b(X, Y) = ( JD X, Y )$. 

In the following section we shall apply this result to realize quadratic Hamiltonians on $\RR^{2n}$ on coadjoint orbits of the Heisenberg Lie group.

\subsection{Quadratic Hamiltonians on coadjoint orbits of the Heisenberg Lie group} In this section we shall realize quadratic Hamiltonians of $\RR^{2n}$ (as in (\ref{ha1})) on coadjoint orbits of the Heisenberg Lie group,
 which are included in a solvable Lie algebra admitting an ad-invariant metric. The corresponding Hamiltonian systems can be written as a Lax pair equation and the solution can  be computed with help of the Adjoint map. Finally we shall prove  involution conditions for a class of functions  on these coadjoint orbits in terms of commutativity on the Lie algebra of derivations of $\hh_n$.

Consider  a linear system of one 
degree  of freedom on $\RR^{2n}$ with Hamiltonian given by:
$$
H(x)= \frac12 (Ax, x)
$$
where   $x=(q_1, \hdots, q_n, p_1, \hdots, p_n)$ is a vector in $\RR^{2n}$ written
  in a symplectic
 basis and $A$ is a symmetric linear operator with respect to the canonical inner product
 $( , )$. This yields the following  Hamiltonian  equation
 
 \vspace{4pt}
 
 $(\ref{ham1}) \qquad \qquad \qquad \qquad \qquad 
 x'=JA x , \qquad\mbox{ with } J = \left( \begin{matrix} 0 & -Id \\ Id & 0
 \end{matrix} \right)
 $
 
 \vspace{4pt}
 
\noindent and being $Id$  the identity.  
The phase space for this system is $\RR^{2n}$. 

We shall construct a solvable Lie algebra that admits an ad-invariant metric on which the system (\ref{ham1}) can be realized as a Hamiltonian system on coadjoint orbits. Moreover it can be written as a Lax pair equation.

Assume that $A$ is non singular and let $b$ denote the bilinear form on $\RR^{2n}=span \{X_i, Y_j\}_{i,j=1}^n$ given by $b(X,Y)=(AX,Y)$.  According to the previous section, $JA$ is skew symmetric with respect to $b$, where $J$ is the canonical complex structure on $\RR^{2n}$ as above. Let $\ggo$ denote the double extension of $(\RR^{2n}, b)$ by $(\RR, JA)$, that is $\ggo=\RR X_0  \oplus \vv \oplus \RR X_{n+1}$ with $\vv= \RR^{2n}$, where the Lie bracket is given by the non trivial relations
\begin{equation}\label{lb}
[U, V] = b(JA U, V) Z_0 \qquad \ [X_{n+1}, U ] = JA U \quad\mbox{ for all } U\in \vv, 
\end{equation}
which can be equipped with the ad-invariant metric defined by
\begin{equation}\label{metric}
\la x_0^1X_0 + U^1 + x_{n+1}^1 X_{n+1}, x_0^2X_0 + U^2 + x_{n+1}^2 X_{n+1}\ra = b(U^1, U^2) + ( x_0^1 x_{n+1}^2 
+ x_0^2 x_{n+1}^1).
\end{equation}

Thus if  $\la , \ra_{\vv}$ denotes the restriction of the metric of $\ggo$ to $\vv=span\{X_i, Y_j\}_{i,j=1, \hdots , n}$, then it coincides with  the non degenerate symmetric bilinear map $b$  of $\RR^{2n}$ and $\ggo$ admits  a orthogonal splitting $ \ggo= \vv \oplus span\{X_0,X_{n+1}\}$. 

Let  $\ggo_{\pm}$ denote the Lie subalgebras 
$$\ggo_+=\RR X_{n+1}, \qquad \ggo_-=\RR X_0 \oplus span\{X_i,, Y_i\}.$$
Clearly they allow the  splitting of $\ggo$ into a vector space direct sum 
  $\ggo = \ggo_+ \oplus \ggo_-$, which via the ad-invariant metric induces the following decomposition $\ggo= \ggo_+^{\perp} \oplus \ggo_-^{\perp}$, direct sum as vector spaces, where
  $$ \ggo_-^{\perp} = \RR \{X_0\} \qquad \qquad \ggo_+^{\perp} = span\{X_i, Y_i\}_{i=1, \hdots, n}
\oplus \RR X_{n+1}.$$
   Indeed   $\ggo_-$ is  an ideal of $\ggo$ isomorphic to the 2n+1-dimensional
Heisenberg Lie  algebra $\hh_n$.

If $G$ denotes a Lie group with Lie algebra $\ggo$, set $G_-\subset G$
the Lie subgroup with Lie subalgebra $\ggo_-$. 
Then  $G_-$ acts on  $\ggo_+^{\perp}$ by the coadjoint
action  $$g_- \cdot X= \pi_{\ggo_+^{\perp}}(\Ad(g_-)X ) \quad g_-\in G_-, \quad X \in
\ggo_+^{\perp},$$
 where $\pi_{\ggo_+^{\perp}}$ denotes the projection of $\ggo$ on
 $\ggo_+^{\perp}$, which 
 in infinitesimal terms gives the following  action of $\ggo_-$ on $\ggo_+^{\perp}$ 
\begin{equation}\label{m22}
\begin{array}{rcl}
\ad^{\ast}_{U} V : = U \cdot V   & =  &   x_{n+1}(V) JA X_{\vv}(U)
\end{array} \qquad \mbox{ for } U\in \ggo_-, \, V\in \ggo_+^{\perp}.
\end{equation}
where $X_{\vv}(U)$ denotes the projection of $U$ onto $\vv$ with respect to the orthogonal splitting 
$\ggo = (\RR X_0 \oplus \RR X_{n+1}) \oplus \vv$. The orbits are 2n-dimensional
if $x_{n+1}(V) \ne 0$  and furthermore $V$ and $W$ belong to the same orbit if and only if
$x_{n+1}(V)=x_{n+1}(W)$, and this allows to  parametrize the orbits by the $x_{n+1}$-coordinate;  
 so we denote them by $\mathcal M_{x_{n+1}}$. The orbits are topologically
like $\RR^{2n}$ since they are 
diffeomorphic to the quotient $\HH_n/Z(\HH_n)$, where  $Z(\HH_n) = \RR X_0$.

Endow the orbits with the canonical symplectic structure of the coadjoint
orbits, that is 
$$\omega_X(\tilde{U_-}, \tilde{V_-})=\la X, [U_-, V_-]\ra= x_{n+1}(X) b(JA U_{\vv}, V_{\vv})$$ for
$X\in \ggo_+^{\perp}$, $U_-, V_-\in \ggo_-$.

Let $f:\ggo \to \RR$ be the ad-invariant function given by $f(X)= \frac12 \la X, X\ra$. The gradient of the function $f$ at a point $X$  is the position vector 
$\nabla f(X) = X$. Since $f$ is
ad-invariant the 
Hamiltonian system of $H=f_{|_{\mathcal M_{x_{n+1}}}}$, the restriction of $f$  
 to the orbit $\mathcal M_{x_{n+1}}$, is given by (\ref{e5}), so we have
\begin{equation}\label{ipe}
\begin{array}{rcl}
\frac{dx}{dt}& = & [\nabla f_+(x),x] = [x_{n+1}X_{n+1},  x_{\vv} + x_{n+1} X_{n+1}]= x_{n+1} JA x_{\vv}\\
 x(0) & = & X^0
\end{array}
\end{equation}
where $X^0 \in \ggo_+^{\perp}$. This Hamiltonian system written as a Lax pair equation is equivalent to  (\ref{ham1}) for $x_{n+1} = x_{n+1}^0 = 1$.  The solution for the initial condition  $X^0 \in \ggo_+^{\perp}$ can be computed with help of the Adjoint map on $G$.  In fact it  can be written as
$$X(t)= \Ad(exp\,\, t x_{n+1}^0 X_{n+1}) X^0.$$
The previous explanations prove the following result.

\begin{thm} Let $H(X)= \frac12(AX,X)$ be a quadratic Hamiltonian on $\RR^{2n}$ with corresponding Hamiltonian system (\ref{ham1}). Then $H$ can be extended to a quadratic function $f$ on a solvable Lie algebra $\ggo$ containing the Heisenberg Lie algebra as a proper ideal. The function $f$ induces  a Hamiltonian system on coadjoint orbits of the Heisenberg Lie group, that can be written as a  Lax pair equation and which is equivalent to (\ref{ham1}). Moreover the trajectories on $\RR^{2n}$ for the initial condition $V^0$ can be computed with help of the Adjoint map on $\ggo$. Explicitely  they are the curves $x(t) = \exp^{t J A} V^0$, where $\exp$ denotes the usual exponential map  of matrices.
\end{thm}

\begin{remark} The Lie algebra $\ggo$ above is isomorphic to the Lie algebra of real functions on $\RR^{2n}$ under Poisson bracket generated by $q_i$, $p_j$, 1 and the Hamiltonian $H$.
\end{remark}

\begin{example}[The motion of n-uncoupled inverse pendula] As example of the previous construction consider  the linear approximation 
of the motion of n uncoupled inverse pendula. This corresponds to the Hamiltonian $
H(x)= \frac12 (Ax, x)$ where  
$$A=\left( \begin{matrix}  Id & 0 \\ 0 & -Id
\end{matrix} \right).
$$
This yields  the Hamiltonian system $x' = JAx$, which in coordinates takes the form
\begin{equation}\label{33}
\begin{array}{rclcl}
\frac{dq_i}{dt} &  = & p_i\\
 \frac{dp_i}{dt} & = & {q_i}
\end{array}
\end{equation}
 
The phase space for this system is $\RR^{2n}$. By considering the setting above we construct coadjoint orbits $\mathcal M$ of the Heisenberg Lie group,  that are included in a solvable Lie algebra $\ggo$ with Lie bracket (\ref{lb}) and ad-invariant metric (\ref{metric}). The Hamilonian system for the restriction to the orbits of the ad-invariant function on $\ggo$ induced by the metric  can be written as 
\begin{equation}\label{ipel}
\begin{array}{rcl}
\frac{dx}{dt}& = & [x_{n+1}X_{n+1},  x_{\vv} + x_{n+1} X_{n+1}]\\
 x(0) & = & X^0
\end{array}
\end{equation}
where $X^0= \sum_i (x_i^0 X_i + y_i^0 Y_i)+x_{n+1}^0 X_{n+1}$. This is in fact a Lax pair equation. If we identify the coordinates $q_i$ with $x_i$ and $p_i$ with $y_i$ then the Hamiltonian system above on the coadjoint orbit $\mathcal M_1$  written in coordinates is clearly equivalent to (\ref{33}). 

The trajectories on $\ggo_+^{\perp}$, $x= \sum_i (x_i(t)X_i+ y_i(t) Y_i) + x_{n+1} X_{n+1}$ are 
parametrized  by

  $$
\begin{array}{rcl}
x_i(t) & = &  x_i^0 \cosh(x_{n+1}^0 t) + y_i^0 \sinh( x_{n+1}^0 t)\\
y_i(t) & = &  x_i^0 \sinh(x_{n+1}^0 t) + y^0_i \cosh(x_{n+1}^0 t)\\
x_{n+1}(t) & = &  x_{n+1}^0
\end{array}
$$

One can verify that the flow at the point $X^0 \in \ggo_+^{\perp}$ is then
\begin{equation}\label{iflow}
\begin{array}{rcl}
\Delta^t(X^0) & = & \sum_i[( x_i^0 \cosh(x_{n+1}^0 t) - y_i^0 \sinh( x_{n+1}^0 t)X_i
 + \\
 & & + (x_i^0 \sinh(x_{n+1}^0 t) + y_i^0 \cosh(x_{n+1}^0 t)Y_i] +  
  x_{n+1}^0 X_{n+1}
  \end{array}
\end{equation}

\vskip .2cm

System (\ref{ipel}) is a Lax pair equation $L^{\prime} = [M, L]= ML -
LM$, taking $L$ and $M$ the following matrices in $M(2n+2,\RR)$ with $\omega_i=1$:
$$ M = \left(
\begin{matrix}
0 & x_{n+1}\omega_1 & 0& 0 &&& &0 & 0\\ x_{n+1}\omega_1 & 0 & 0 & 0 
&&&&0 &
0 \\ 0 & 0 & 0 & x_{n+1}\omega_2 &&&&0& 0\\ 0& 0 & x_{n+1}\omega_2&
0&&&&0&0\\ & & & & \ddots & & &\vdots & \vdots\\ & & & & & 0 &
x_{n+1} \omega_n&0 & 0\\ & & & & 0& x_{n+1}\omega_n& 0 &0& 0\\ 0& 0& 
\hdots
& & & & & 0 & 0\\ 0 & 0 & \hdots & & & &  & 0 & 0
\end{matrix}
\right) $$
$$ L = \left(
\begin{matrix}
0 & x_{n+1}\omega_1 & 0& 0 & & & & &x_1\\ x_{n+1}\omega_1 & 0 & 0 & 0  
& &
& & & y_1\\ 0 & 0 & 0 & x_{n+1}\omega_2 & & & & & x_2\\ 0& 0 &
x_{n+1}\omega_2 & 0 & & & & & y_2\\ & & & & \ddots & &  & \vdots &
\vdots\\ & & & & & &  x_{n+1} \omega_n& 0& x_n\\ & & & & &  x_{n+1}
\omega_n & 0 &0 & y_n\\ -\frac12 y_1& \frac12 x_1&-\frac12 y_2 & \frac12
x_2 &\hdots & -\frac12 y_n& \frac12 x_n &  0 & 0\\ 0& 0 & 0 & 0 &
\hdots  & 0& 0& 0 & 0
\end{matrix}
\right) $$

Next we shall prove the complete integrability (\ref{ci}) of the function $H$  with system (\ref{ipel}). We shall show a
set of n-functions that are in involution. As in the case of the n-uncoupled harmonic oscillators they are not Ad-invariant,
hence they do not satisfy conditions of Theorem (\ref{AKS1}) and Theorem
(\ref{AKS2}) does not hold in this case.

 Consider $H_i={f_i}_{|_{\mathcal M_{x_{n+1}}}}$ be the 
restrictions to a orbit $\mathcal M_{x_{n+1}}$ of the
functions
\begin{equation}\label{fip}
f_i(X)=  \frac12 (y_i^2-x_i^2) \quad \mbox{ for }i= 1, \hdots n
\end{equation}
  The
functions $f_i$ are not ad-invariant but their restrictions Poisson commute on the orbit. Moreover we assert

\begin{prop} 
The function $H$ is completely integrable (in the sense of (\ref{ci})) 
on the orbits $\mathcal
M_{x_{n+1}}$  for all $x_{n+1}\neq 0$. 
\end{prop}
 \begin{proof} The proof follows from a analogous procedure as in (\ref{ci1}).
\end{proof}

However in this case the set
$$\begin{array}{rcl}
\mathcal N_{x_{n+1}} & = & \cap_{i=1}^n \{X \in \mathcal M_{x_{n+1}} : H_i(X) =
c_i\}\\
& = & \cap_{i=1}^n\{X= \sum_i(x_i X_i + y_i Y_i) +x_{n+1} X_{n+1}\in 
\mathcal M_{x_{n+1}}
: x_i^2+ y_i^2 = c_i\}
\end{array}$$
 is not compact.
\end{example}

\begin{remark} If $\ggo_{(4)}$ denotes the four dimensional Lie algebra of the example above, the connected Lie group $G=\exp \ggo$ is called the Boidol group.
\end{remark}

Motivated by the involution conditions proved in the equation of motion of both systems corresponding to n-uncoupled harmonic oscillators and n-uncoupled inverse pendula, we shall investigate involution conditions on the coadjoint orbits of the Heisenberg Lie group for the restrictions of the quadratic functions $f(X)=\frac12 \la X,X\ra$, where $\la , \ra$ denotes the ad-invariant metric on the solvable Lie algebra $\ggo$. 

Let $g_i, g_j$ be two quadratic on $\RR^{2n}$  associated to symmetric 
maps $A_i, A_j:\vv \to \vv$ respectively, that is
$$g_i(X) = \frac12 (A_i X, X) \qquad \qquad g_j(X)=\frac12 (A_jX,X).$$ 
Consider the quadratic functions on the solvable Lie algebra $\ggo$, that are extensions of $g_i, g_j$ to $\RR X_0 \oplus \RR X_{n+1}$, for instance as 
$$g_i(X) = \frac12 (A_i X_{\vv}, X_{\vv}) + x_0 x_{n+1}\qquad \qquad g_j(X)=\frac12 (A_jX_{\vv},X_{\vv}) +x_0 x_{n+1}.$$
For the following results these extensions are not unique. For instance extending them trivially we get the same conclusions.

 Let $H_i, H_j$ denote the restrictions
of $g_i, g_j$ to the orbits $\mathcal M_{x_{n+1}}$ and let  $X\in \mathcal M_{x_{n+1}} \subset \ggo_+^{\perp}$. The Poisson bracket of 
the functions $H_i, H_j$ on the orbit follows:  
$$
\{H_i, H_j\}(X)  =  \la X, [\nabla {g_i}_-(X), \nabla {g_j}_-(X)]\ra 
$$
Thus we need to compute the gradients of $g_i, g_j$, which are
$$\nabla g_i(X) = A^{-1} A_i X_{\vv}+x_0X_0 + x_{n+1} X_{n+1} \qquad \nabla g_j(X) = A^{-1} A_j X_{\vv}+x_0 X_0 + x_{n+1} X_{n+1}.$$
In Theorem (\ref{bij}) we established a correspondence between the symmetric maps on $\RR^{2n}$ and the derivations of $\hh_n$ acting trivially on the center, given by $\psi$ which sends $A \to JA$, where $J$ is the canonical complex structure on $\vv$ and $JA$ is extended trivially to the center of $\hh_n$.

\begin{thm} \label{c11} The functions $H_i, H_j$ are in involution on the orbits $\mathcal
M_{x_{n+1}}$ if and only if 
\begin{equation}\label{c1}
[\psi(A_i), \psi(A_j)]=0
\end{equation}
\end{thm}
\begin{proof} Let $X\in \mathcal M_{x_{n+1}} \subset \ggo_+^{\perp}$. Then for
the functions $H_i, H_j$ the Poisson bracket on the orbit follows:
$$\begin{array}{rcl}
\{H_i, H_j\}(X) & = & \la X, [A_i X_{\vv}, A_jX_{\vv}]\ra = \la x_{n+1} [X_{n+1},
A^{-1}A_i X_{\vv}], A^{-1}A_j X_{\vv}\ra \\
& = &  x_{n+1} \la J A_i X_{\vv}, A^{-1} A_jX_{\vv}\ra = x_{n+1} (J A_i X_{\vv}, A_jX_{\vv})
\end{array}
$$
  Therefore $\{H_i, H_j \}(X)=0$ if and only if
$( A_j J A_i X_{\vv}, X_{\vv}\ra=0$ which is equivalent to  
$A_j J  A_i= A_i J  A_j$, if and only if $JA_j J  A_i= JA_i J  A_j$. Since $\psi(A) = JA$ we proved the result.
\end{proof}


\begin{cor} If there exists an n-dimensional abelian subalgebra on $z(JA)_{\dd}$, where 
$$z(JA)_{\dd}=\{ D \in \dd \mbox{ such that } [D, JA]=0\}$$
 then the Hamiltonian function $H$ restriction of the function $f(X)= \frac12(AX,X)$ is completely integrable on the orbits $\mathcal M_{x_{n+1}}$ for $x_{n+1} \neq 0$.
\end{cor}
\begin{proof} The previous theorem says that  the restrictions to the orbit $\mathcal M_{x_{n+1}}$ of the functions $g_i, g_j$ are in involution if their corresponding derivations commute in $\dd$. In particular for $g_i$ and $f$, we have that $H$ and $H_i$ Poisson commute on the orbit if and only if $JA_i$ belongs to the centralizer of $JA$ in $\dd$, $z(JA)_{\dd}$. Since the complete integrability requires of n linearly independent functions, this can be done with a basis of a $n$-dimensional abelian subalgebra of $z(JA)_{\dd}$, finishing the proof.
\end{proof}

Thus using Lie Theory we can say that many of these systems are completely integrable. In fact the Lie algebra of derivations of $\hh_n$ acting trivially on the center is $sp(n)$ (see \cite{Sa}), which  has a Cartan decomposition o the form $sp(n)=\uu \oplus \pp$, where $\uu$ is a maximal compact subalgebra.  Taking for instance  any element in $\uu$, then it induces a completely integrable system. More generally we should study the abelian  subalgebras on $sp(n)$.
 
\begin{example} [Involution on the oscillators] Note that whenever  we choose $A=Id$  as the symmetric map for $H$ in
(\ref{ha1}),  and we apply the double extension procedure we get the oscillator Lie algebra $\ggo$ endowed with its Lorentzian ad-invariant metric.  Let $H$ denote as above the restriction to the coadjoint orbits of the function induced by the metric. Reading the previous proposition in this situation we have that:

\vspace{4pt}

$\{H, H_i\} =0$ if and only if $\psi(A_i)$ belongs to the Lie subalgebra of isometries of $\HH_n$ fixing the identity and acting trivially on the center.

\vspace{4pt}

In fact the Lie algebra of the isometries of $\HH_n$ fixing the identity is the set of skew symmetric derivations of $\hh_n$, that is those derivations satisfying $\la Du, v\ra'=-\la u, Dv\ra'$, where $\la , \ra'$ is the canonical inner product on $\HH_n$ as in (\ref{hn}). The previous results says that $H_i$ Poisson commutes with $H$ if and only if $A_i J = JA_i$. But $JA_i$ can be identified with a derivation $D$ of $\hh_n$ acting trivially on the center and $(JA_i)^t= -A_iJ = -JA_i$, that is $D$ is skew symmetric, and this proves our assertion. 

Explicitely, let $J:\vv \to \vv$ denote the canonical complex structure of $\RR^{2n}$ defined as  (\ref{ham1}) in 
 the Introduction. Notice that by identifying $\RR^{2n}$ with $\vv$ as isometric vector spaces then $J$  coincides with the restriction of $\ad_{X_{n+1}}$
 to $\vv$ in the oscillator Lie algebra $\ggo$. 
 The restriction to the orbit of a quadratic $g(X)=\la AX, X \ra$ with  a 
 symmetric map $A:\vv \to \vv$
 of the form
 $$\left( \begin{matrix} B & C \\ D & E \end{matrix}\right)
 $$
 Poisson commutes with $H$  the restriction of the quadratic induced by the metric on  $\ggo$ 
if and only if $C=-D$ and $B=E$  where $B$ is also symmetric.  That is, the matrix $A$ seen as a linear map on $\RR^{2n}$ is complex and symmetric. 
Assume now that the restrictions of two quadratic functions $g_i, g_j$ associated to symmetric maps
$A_i, A_j$  are in involution with $H$. Then they pairwise
 Poisson commute if and only if $[C_i, B_j]=[C_j, B_i]$ and $[C_i, C_j]=[B_i, B_j]$ for all i,j, where $[ , ]$ is the canonical Lie bracket for matrices: $[A,B]=AB-BA$.

 In particular the functions $H_i$ defined as the restrictions to the orbit
 $\mathcal M_{x_{n+1}}$ of the
 functions $f_i(X)=\frac12(x_i^2+y_i^2)+ x_0 x_{n+1}$ as in (\ref{fi}) are examples of the
 quadratic functions above. 
 \end{example}

\

{\bf Conclusions} We studied Hamiltonian systems on coadjoint orbits of the
Heisenberg Lie algebra. This was done with Lie theory as a powerful tool. The Heisenberg Lie algebra was
included in a solvable Lie algebra admitting an ad-invariant metric. The metric
was useful to write the corresponding Hamiltonian system in a Lax form. Next we study Poisson commutativity. The  involution conditions for a class of functions are related to the Heisenberg Lie algebra. This work shows examples of applications of the Adler-Kostant-Symes scheme on solvable Lie algebras. This allows to ask for new examples and a more general theory for Poisson commutativity.


\begin{thebibliography}{GGGG}


\bibitem[AM]{A-M} {\sc Abraham, R., Marsden, J}, {\it Foundations of 
Mechanics
}, Second edition. The Benjamin Cummings publishing company, (1985).

\bibitem[A]{A} {\sc Adler, M.}, { On a trace for formal 
pseudodifferential
operators and the symplectic structure for the KdV type equations}, 
Invent. Math., {\bf 50},  219-248, \,(1979).

\bibitem[Ar]{Ar} {\sc Arnold, V. I.}, {\it Mathematical methods of 
classical
mechanics}, Springer Verlag, \,(1980).


\bibitem[B-K]{BK} {\sc Baum, H., Kath, I.}, {Doubly extended Lie
  groups ' curvature, holonomy and parallel spinors},
  Diff. Geom. Appl., {\bf 19},  253-280, \,(1998).  

\bibitem[G]{G}{\sc  Guest, M}, {\it Harmonic Maps, Loop Groups and 
Integrable
Systems}. (London Math.Soc.Student Texts; 38). New York: Cambridge 
University
Press (1997).

\bibitem[G-S]{GS}{\sc  Guillemin, V., Sternberg, S.}, {\it Symplectic 
tecniques
in physics}. Cambridge New York Port Chester Melbourne Sydney: 
Cambridge
University
Press (1991).

\bibitem[F-M]{FM} {\sc Fomenko, A., Mischenko, A.}, { Generalized 
Liouville
method of integration of Hamiltonian systems}, Funct. Anal. and its 
Applic., {\bf
12},  113 -121, \,(1978).

\bibitem[Ko1]{Ko1}{\sc Kostant, B.}, { Quantization and Representation 
Theory,
in: {\it Representation Theory of Lie groups, Proc. SRC/LMS Res. Symp., 
Oxford
1977. London Math. Soc. Lecture Notes Series}}, {\bf 34}, 287-316, 
(1979).

\bibitem[Ko2]{Ko2}{\sc Kostant, B.}, { The solution to a generalized 
Toda lattice
and representation theory}, Advances in Math., {\bf 39}, 195 - 338,
(1979).


\bibitem[M]{M} {\sc Medina, A.}, { Groupes de Lie munis de pseudo-metriques de Riemann bi-invariantes}, S\'em. G\'eom. Diff., (1981-1982), Montepellier.


\bibitem[M-R]{MR} {\sc Medina, A., Revoy, Ph.}, { Alg\`ebres de Lie et 
produit
scalaire invariant}, Ann. scient. \'Ec. Norm. Sup., 4$^e$ s\'erie, {\bf 
t. 18},
 391 - 404,\,(1985).

\bibitem[O1]{O1} {\sc Ovando, G.}, {Estructuras complejas y sistemas
hamiltonianos en grupos de Lie solubles}, Tesis Doctoral, Fa.M.A.F. 
Univ. Nac.
de C\'ordoba,( Marzo 2002).


\bibitem[Ra]{Ra} {\sc Raghunathan, M.}, {Discrete subgroups of Lie groups},
Springer, New York,(1972).


\bibitem[R1]{R1} {\sc Ratiu, T.}, {\it Involution theorems}, Geometric 
methods in Math.
Phys., Lect. Notes in Math., {\bf
775}, Procedings, Lowell, Massachusetts 1979, Springer Verlag, (1980).


\bibitem[R2]{R2} {\sc Ratiu, T.}, { The motion of the free 
n-dimensional rigid body},
Indiana Univ. Math. Journal, {\bf 29}, 609 - 629, (1980).


\bibitem[Sa]{Sa} {\sc Saal, L.}  { The automorphism group of a Lie algebra of Heisenberg type} Rend. Sem. Mat. Univ. Pol. Torino, {\bf 54} 2, (1996).

\bibitem[Sy]{Sy}{\sc Symes, W.}, {Systems of Toda type, inverse
spectral
problems and representation theory}, Invent. Math., {\bf 59}, 13 - 53,
(1978).

\bibitem[Va]{Va} {\sc Varadarajan, V.}, {\it  Lie groups, Lie algebras
and
their representations}, Springer, (1984).


\end{thebibliography}
\end{document}